\documentclass[final,3p]{article}
\usepackage[T1]{fontenc}
\usepackage{mathptmx}
\usepackage[top=2.5cm,bottom=2.5cm,left=2.1cm,right=2.1cm,footskip = 1.25cm, includehead, includefoot]{geometry}
\usepackage{amssymb}
\usepackage{enumitem}
\usepackage{natbib}
\usepackage[T1]{fontenc}
\usepackage{tgtermes}

\renewcommand{\bibliography}{REFERENCES}

\usepackage{lineno}
\title{ON THE VESSEL ENERGY REQUIREMENT PREDICTION FROM THE ACCELERATION STAGE TOWING EXPERIMENTS ON MODELS}
\author{Klaudia Wrzask\\ Gdansk University of Technology}
\date{\today}

\begin{document}
	
	\maketitle
	
	\begin{abstract}
		\textit{
			One of the most crucial tasks for naval architects is computing the energy required to meet the ship's operational needs. When predicting a ship's energy requirements, a series of resistance tests on a scaled model vessel is carried out in the constant speed stage. Another important component is the ship's hydrodynamic added mass, which should also be considered when performing the seakeeping analysis. The second law of dynamics states that all this information, that is, the hull resistance dependence on the vessel's speed and the added mass, is accessible from just one acceleration stage towing test done up to the maximal speed. Therefore, this work aims to generalize Froude's scaling procedure from the model to full-scale vessels for accelerated motion. 
		}
		
	\end{abstract}
	
	
	\section*{\centering INTRODUCTION}
	\qquad
	In 1870, W. Froude initiated an investigation into ship resistance with the use of vessel models. The resistance is the horizontal component of the force opposing the forward motion of the vessel's hull. Froude noted that the wave configurations around geometrically similar forms were similar if compared at speeds proportional to the square root of the model length. He propounded that the total resistance could be divided into skin friction resistance and residuary, mainly wave-making resistance. Specific residuary resistance would remain constant at corresponding speeds between the model and the full-scale vessel. Next, estimates of frictional resistance from a series of measurements on planks of different lengths and with different surface finishes were derived, \cite{Froude1872}. The scaling procedure proposed by Froude is based on constant speed towing experiments on the model vessel. Further, \cite{Froude1874} carried out full-scale tests on HMS Greyhound (100 ft), and the results showed substantial agreement with the model predictions. Finally, \cite{Froude1877} gave a detailed explanation of wave-making resistance, supporting his scaling methodology. Froude's ideas still dominate this subject.
	
	Nowadays, when predicting ships' energy requirements, resistance tests on model vessels are still conducted. For constant speed, the resistance is determined by towing force measurements. In the next step, the resistance test results are scaled from the model to the full-scale ship. A modification of Froude's scaling method by spitting the residual resistance into the form resistance, and the wave resistance, suggested by \cite{{Hughes1954}}, and known as the form factor $(1+k)$ approach, was later adopted by the International Towing Tank Conference (ITTC).
	
	The scaling procedures, as mentioned above, refer to constant speed towing tests. Another important aspect is the derivation of the hydrodynamics added mass of the ship, which may account for up to 30\% of the ship's mass and represents significant inertia for the accelerated motion. It follows from the laws of dynamics that all this information, i.e., the resistance dependence of speed and the added mass, is accessible from the acceleration stage towing test done up to the maximum speed. The measuring apparatus in the 19th century did not allow Froude to conduct his research on the acceleration stage with the same level of precision as it is possible now. Despite the development, great precision, and sampling rate of measurements, the author is unaware of any scaling procedures from the acceleration stage towing tests. Therefore, this paper derives a dynamical scaling proposition for the propulsion force needed to estimate the full-scale ship's energy consumption. A fully dynamical model can simulate any profile: constant speed, acceleration, deceleration, and gliding. Moreover, such an approach allows to optimize of the required energy based on the dynamical systems, especially desired for short-range vessels, where constant speed is not the major stage, \cite{Kunicka2019}. 
	
	The organization of this paper is as fallows. First all the components of the ship's  resistance in accelerated motion are going to be introduced. Next, the scaling procedure for constant speed towing test is going to be explained. Further, a proposition for the towing tests in accelerated stage and a scaling procedure are explained.

	\section*{\centering NEWTON'S SECOND LAW OF DYNAMICS FOR TOWING TESTS}\label{sec:Newton}
	\qquad
	To explain the concept of the scaling procedure for any acceleration, let us start with Newton's second law of dynamics, which for any vessel takes the following form:
	\begin{eqnarray}
		m \textrm{v}'&=&F_P(\textrm{v},\textrm{v}')-R_T(\textrm{v}).
		\label{secondLaw2}
	\end{eqnarray}
	Here $m$ stands for the total mass, which is the sum of the mass of the vessel $m_v$ and the hydrodynamic added mass of the water $m_{add}$, i.e., 
	\begin{eqnarray}
		m=m_v+m_{add}.
	\end{eqnarray}
	In general, the added mass is a second-order tensor relating the fluid acceleration vector to the resulting force vector on the body. Only the surge added mass is taken under consideration in this work. Further, $\textrm{v} '$ denotes the speed derivative over time, $F_P$ is the propulsion force for the full-scale vessel, or the towing force in the case of the model vessel, and $R_T$ is the total hull resistance force. 
	
	In the case of constant speed, i.e., $\textrm{v}'=0$, the towing force is equal in magnitude to the total hull resistance, and the second law takes the following form
	\begin{eqnarray}
		F_P(\textrm{v},0)&=&R_T(\textrm{v}).
	\end{eqnarray}
	Then, after rewriting (\ref{secondLaw2}) we get
	\begin{eqnarray}
		F_P(\textrm{v},\textrm{v}')&=&F_P(\textrm{v},0)+(m_v+m_{add}) \textrm{v}'.
		\label{secondLawacc}
	\end{eqnarray}
	Formula (\ref{secondLawacc}) shows that from the acceleration stage towing tests, which give data $F_P(\textrm{v},\textrm{v}')$, information on both the total hull resistance dependence on constant speed $F_P(\textrm{v},0)$ and the added mass $m_{add}$ are accessible.
	\subsection*{THE ADDED MASS}\label{sec:}
	\qquad
	In 1786 Du Buat found by experiment that the motion of spheres oscillating in water could only be described if an added mass was included in the equations of motion. In fluid mechanics, the added mass is defined as an extra fluid mass that accelerates with the body. It is the inertia added to a system because the accelerating body, to pass through, must move aside and then close in behind a specific volume of the surrounding fluid. The fluid thus possesses kinetic energy that it would lack if the immersed body were not in accelerated motion. The body has to impart this kinetic energy to the fluid by doing work on the fluid. Any corresponding equations of motion for the immersed body must take into account this loss of kinetic energy. This can be modeled in the equations of motion as some volume of fluid moving with the object, although, in reality, the fluid will be accelerated to varying degrees. When the body moves at a constant speed, the corresponding fluid's motion is steady; thus, the kinetic energy of the fluid is constant. It follows that for constant-speed motion, the added mass terms can be omitted in the equations of motion, \cite{Imlay1961}. 
	
	Extra resistance caused by the accelerated objects in a fluid medium has been investigated both experimentally and theoretically: \cite{Poisson1832}, \cite{IversenBalent1951}. The findings are that the added mass depends on the size and shape of the immersed body, the direction in which it moves through the fluid with respect to its axis, and the density and viscosity of the fluid. 
	
	The added mass can be described by a dimensionless coefficient which depends on the shape of the immersed body. The dimensionless added mass coefficient $C_M$ is the added mass divided by the displaced fluid mass, \cite{Newman2018}; that is, divided by the fluid density $\rho$ times the volume of the body under water $V$, and therefore
	\begin{eqnarray}
		m_{add}=C_M \rho V.
	\end{eqnarray}
	
	The same principles apply to ships. In the marine sector, added mass is referred to as hydrodynamic added mass. The hydrodynamic added mass has also been investigated in the maritime area. \cite{Motora1960} first conducted model testing for a ship called Mariner to predict the added mass. \cite{Ghassemi2011} proposed a numerical calculation of the marine propeller's added mass using Boundary Element Method. \cite{Zeraatgar2020} investigated the surge added mass of planing hulls by model vessel experiments and by approximates with a  quasi-analytical method. The conclusion was that the surge added water mass could account for 10\% of the total mass for the investigated planing hulls. 
	
	Essentially for ships, the added mass can reach even one-third of their mass, representing significant inertia in addition to viscous and wave-making drag forces. Thus, the energy required to accelerate the added mass should also be considered when performing a seakeeping analysis. 
	
	When conducting a towing test in the acceleration stage, the surge added mass can be obtained from the equations of motion by extrapolating the towing force to zero speed $F_P(0,\textrm{v}')$, i.e.,
	\begin{eqnarray}
		(m_v+m_{add}) \textrm{v}'&=&F_P(0,\textrm{v}')-R_T(0).
	\end{eqnarray}
	Then the total hull resistance can be neglected $R_T(0)=0$, and it follows that
	\begin{eqnarray}
		m_{add}&=&\frac{F_P(0,\textrm{v}')}{\textrm{v}'}-m_v.
		\label{addM}
	\end{eqnarray}

	\subsection*{THE TOTAL HULL RESISTANCE }\label{sec:}
	\qquad
	Even in calm water, the ship experiences water's resistance to its motion. This force is referred to as the total hull resistance $R_T$. This resistance force is needed to calculate the ship's effective power. Many factors combine to form the total resistance force acting on the hull. The physical factors affecting ship resistance are the friction and viscous effects of water acting on the hull and the energy required to create and maintain the ship's characteristic bow and stern waves. Finally, the minor contribution has the resistance that the air provides to the ship's motion. This may be written in the following form:
	\begin{equation}
		R_T=R_V+R_W+R_A,
	\end{equation}
	where: 
	$R_T$ is the total hull resistance, 
	$R_V$ is the viscous friction resistance,  
	$R_W$ is the wave-making resistance,  and 
	$R_A$ stands for the air resistance. 
	
	The total hull resistance $R_T$ can also be formulated by means of the dimensionless coefficient $C_T$ with the following equation:
	\begin{equation}
		R_T=\frac{1}{2}C_T\rho S \textrm{v}^2.\label{RT}
	\end{equation}
	Here: 
	$\rho$ is the water density, 
	$S$ is the wetted surface area of the underwater hull, and 
	$\textrm{v}$ is the speed of the vessel.
	
	As the total hull resistance $R_T$ is the sum of viscous $R_V$ and wave-making $R_W$ resistance, neglecting the air resistance, one can write an equation for the total dimensionless resistance coefficient in terms of the viscous and wave-making coefficients, such that
	\begin{equation}
		C_T = C_V + C_W,
	\end{equation}	
	where: 
	$C_T$ is the coefficient of the total hull resistance, 
	$C_V$ is the coefficient of the viscous frictional resistance, and 
	$C_W$ is the wave-making resistance coefficient. 
	
	To quantify these dimensionless resistance coefficients, two numbers are used. The Reynolds number ${\mathcal Re}$ quantifies the influence of viscous forces on the fluid's motion. It indicates the ratio of inertial to viscous forces and, for the ship, is defined as a dimensionless ratio
	\begin{equation}
		{\mathcal Re}=\frac{\textrm{v} \rho L}{\mu}=\frac{\textrm{v} L}{\nu},
	\end{equation}
	where: 
	$\textrm{v}$ is the vessels speed, 
	$L$ is the length of the wetted surface,
	$\mu$ is the dynamic viscosity, and $\nu$ is the kinematic viscosity. 
	
	The Froude number ${\mathcal Fr}$, in hydrology and fluid mechanics, is used to quantify gravity's influence on fluid's motion. It indicates the ratio of the inertia forces to the gravitational forces related to the mass of water displaced by a floating vessel. It is defined by a dimensionless ratio:
	\begin{equation}
		{\mathcal Fr}=\frac{\textrm{v}}{\sqrt{g L}}.
	\end{equation}
	Here $g$ denotes the gravity acceleration. Then, the relationship between these two numbers can be written in the following, practical for scaling purposes, form
	\begin{equation}
		{\mathcal Re}=\frac{\rho}{\mu}\sqrt{g}L^{1.5}{\mathcal Fr}.
	\end{equation}
	
	\subsection*{THE VISCOUS RESISTANCE }\label{sec:}
	\qquad
	Although water has low viscosity, it produces a significant friction force opposing the ship's motion. For the viscous resistance $R_V$, the skin friction resistance and the viscous pressure resistance contribute. Experimental data have shown that water friction can account for most of the hull's total resistance at low speeds and is still dominant for higher speeds, \cite{Birk2019}. A ship's hull shape can influence the viscous pressure drag's magnitude. Vessels with a lower length-to-beam ratio will have greater drag than those with a larger length-to-beam ratio. 
	
	The dimensionless viscous coefficient $C_V$, taking into account both the skin friction and the viscous pressure resistance, can be derived from the formula:
	\begin{equation}
		C_V = (1+k)C_F. \label{CV}
	\end{equation}
	Here $(1 + k)$ is the form factor that depends on the hull form, and $C_F$ is the skin friction coefficient based on flat plate results. The form factor $(1 + k)$ can be derived from low-speed tests when, at low Froude numbers ${\mathcal Fr}$, the wave resistance coefficient $C_W$ tends to zero and therefore $(1 + k) = C_T/C_F$. The skin friction resistance coefficient $C_F$ is assumed to be dependent on the Reynolds number ${\mathcal Re}$ and is recommended to be calculated through the ITTC-1957 skin friction line as
	\begin{equation}
		C_F({\mathcal Re})=\frac{0.075}{(\log_{10} {\mathcal Re}-2)^2}.\label{CF}
	\end{equation}
	The ITTC-1978 powering prediction procedure for deriving the viscous coefficient $C_V$ recommends the use of formula (\ref{CF}), together with the form factor $(1+k)$. 
	\subsection*{THE WAVE-MAKING RESISTANCE }\label{sec:}
	\qquad
	When a submerged vessel travels through a fluid, pressure variations are created around the body. Near a free surface, the pressure variations manifest themselves through changes in the fluid level, creating waves. Such a wave system is made up of transverse and divergent waves. With a body moving through a stationary fluid, the waves travel at the same speed as the body. It follows that the transverse wavelength depends then on the speed of the ship. The mathematical form of such a wave system is called the Kelvin wave after Lord  \cite{Thomson1887}.	The first step for formulating an analytical expression for the wave resistance was taken by \cite{Michell1898}. A review of Michell's approach and its impact on ship hydrodynamics is given by \cite{Tuck1989}. Further wave resistance was investigated both theoretically and experimentally by \cite{Havelock1909} and elaborated in\cite{Havelock1932}. The findings are that the amplitudes of the waves directly depend on the ship's Froude number ${\mathcal Fr}$. Thus the dimensionless coefficient for the wave-making resistance $C_W$ is assumed to be dependent only on the Froude number. The wave resistance for low speeds is negligible, but for Froude numbers over 0.35, wave resistance may exceed the viscous resistance for most vessels, \cite{Birk2019}. Setting equal Froude numbers for the model and full-scale vessel, such that the wave resistance coefficients are equal, still dominates the subject of scaling procedures. This assumption will also be used in the proposed scaling procedure for the acceleration stage towing tests.

	\subsection*{THE RESISTANCE BREAKDOWN }\label{sec:}
	\qquad
	Within the subject of the resistance breakdown, it is worth emphasizing the fundamental difference between the scaling methods proposed by Froude and Hughes. Froude assumed that all residuary resistance scales according to Froude's law, that is, for the same Froude number ${\mathcal Fr}$. This is not physically correct because the viscous pressure drag included within $C_V$ dimensionless coefficient should scale according to Reynold's law. Hughes assumes that the total viscous resistance, i.e., the friction and the form, scales according to Reynold's law. This leads to the dimensionless resistance coefficient breakdown:
	\begin{equation}
		C_T({\mathcal Re}, {\mathcal Fr})=C_V({\mathcal Re})+C_W({\mathcal Fr}).\label{breakdown}
	\end{equation}
	This also needs to be adjusted, as the viscous resistance interferes with the wave-making resistance. The reason is that the boundary layer growth suppresses the stern wave; thus, the wave resistance can depend on ${\mathcal Re}$. Moreover, the viscous resistance depends on the pressure distribution around the hull, which depends on wave-making, \cite{Molland2017}. Thus, an interaction term $C_{INT}({\mathcal Re},{\mathcal Fr})$, depending on both numbers is non-zero, i.e.,
	\begin{equation}
		C_T({\mathcal Re}, {\mathcal Fr})=C_V({\mathcal Re})+C_W({\mathcal Fr})+C_{INT}({\mathcal Re},{\mathcal Fr}).\label{interaction}
	\end{equation}
	Therefore, the resistance breakdown is an assumption for the scaling practice rather than an exact physical representation. A detailed outline of the scaling effects and evidence supporting the existence of an interaction term is given by \cite{Terziev2022}. Nevertheless, the overall error caused by the resistance coefficient breakdown assumption (\ref{breakdown}) is sufficiently small. The form factor method proposed by Hughes and adopted by the ITTC is still an extremely valuable tool in predicting ships' energy requirements.
	
	For the dynamical scaling purpose of this paper, assumptions, as mentioned above, will also be made; that is, the viscous friction coefficient $C_V$ depends only on the Reynolds number ${\mathcal Re}$, the wave-making coefficient $C_W$, only on the Froude number ${\mathcal Fr}$, and the interaction term will be neglected.

	\section*{\centering SCALING PROCEDURE FOR CONSTANT SPEED TOWING TESTS}\label{sec:scalingConstantSpeed}
	\qquad
	Before explaining the scaling procedure from the acceleration stage, let us look at the constant speed stage towing tests because most assumptions will be the same for both approaches.
	To perform a scaling procedure for constant speeds, first, a geometric scale $\lambda$ is set as the ratio of the full-scale ship length $L_S$ to the model vessel length  $L_M$, i.e.,
	\begin{equation}
		\lambda=\frac{L_S}{L_M}.
	\end{equation}
	Then for equal Froude numbers of both the full-scale ship and the model vessel: ${\mathcal Fr}_M={\mathcal Fr}_S$, the Froude's law of similarity sets the corresponding speeds:
	\begin{equation}
		\frac{\textrm{v}_S}{\textrm{v}_M}=\lambda^{0.5}.
		\label{similarity}
	\end{equation}
	Here: 
	$\textrm{v}_S$ is the full-scale ship speed and  
	$\textrm{v}_M$ denotes the model vessel speed. Newton's second law of dynamics (\ref{secondLaw2}) for constant speeds takes the following form for both the full-scale and the model vessel
	\begin{eqnarray}
		0&=&F_P(\textrm{v},0)-R_T(\textrm{v}).
		\label{secondLawConstantSpeed}
	\end{eqnarray}
	Therefore, for constant vessel speeds, the towing force is equal in magnitude to the total hull resistance force, thus
	\begin{eqnarray}
		F_P(\textrm{v},0)=R_T(\textrm{v}).
		\label{FforContsSpeed}
	\end{eqnarray}
	Moreover, the propulsion force needed to asses energy requirement for constant full-size vessel speeds is equal to the total resistance force acting on the full-size hull.
	In general, the scaling procedure for determining the total hull resistance of a full-scale ship from constant speed towing experiments on a geometrically scaled model vessel may be described in the following steps:
	\begin{enumerate}[label=Step \arabic*:, leftmargin=*,itemsep=0mm]
		\item Setting the full-scale ship speed $\textrm{v}_S$ range, from the minimum to the desired maximum ship speed.
		\item Calculating the corresponding towing speeds for the model $\textrm{v}_M$ using the Froude's law of similarity (\ref{similarity}).
		\item Recording, from the constant speed stage, the total hull resistance force $R_T(\textrm{v}_M)$ of the model vessel towed in a series of tests at each speed $\textrm{v}_M$.
		\item Determining the coefficient of the total hull resistance for the model at each speed $C_T(\textrm{v}_M)$ from formula (\ref{RT}).
		\item Determining the coefficient of the viscous resistance for the model vessel at each speed $C_V(\textrm{v}_M)$ using the ITTC recommended formulas (\ref{CV}) and (\ref{CF}).
		\item Calculating the wave-making coefficient for the model vessel at each speed  $C_{W}(\textrm{v}_M)=C_T(\textrm{v}_M)-C_V(\textrm{v}_M)$.
		\item The wave-making resistance coefficients for the full-scale and the model vessel are equal: $C_{W}(\textrm{v}_S) = C_{W}(\textrm{v}_M)$.
		\item
		Determining the coefficient of the viscous resistance for the full-scale ship $C_{V}(\textrm{v}_S)$, at speeds corresponding to the model towing speeds, with the use of the ITTC recommended formulas (\ref{CV}) and (\ref{CF}).
		\item 
		Calculating the dimensionless coefficient of the total hull resistance for the full-scale vessel at each speed $C_{T}(\textrm{v}_S) = C_{W}(\textrm{v}_S)+C_{V}(\textrm{v}_S)$.
		\item Determining the total hull resistance of the full-scale vessel for each speed using formula (\ref{RT}). 
	\end{enumerate}
	\section*{\centering PROPOSED SCALING PROCEDURE FROM THE ACCELERATION STAGE TOWING TESTS}\label{sec:}
	\qquad
	The proposed scaling procedure for accelerated motion has the same methodology as Froude's scaling for constant speed. The difference is that we are going to take a step back with the equation of motion (\ref{secondLawConstantSpeed}) to full dynamics (\ref{secondLaw2}) because in accelerated motion, the towing force is needed to overcome the total hull resistance and accelerate the model vessel. 
	
	Table \ref{scaling} presents basic assumptions for the scaling rules needed in accelerated motion. Lower indexes $_S$ correspond to the full-size vessel and $_M$ to the model vessel.
	\begin{table}[h!]
		\centering
		\caption{scaling rules and basic assumptions}
		\vspace{6pt}
		\begin{tabular}{lll} 
			\hline
			Physical quantity &Scaling rule& Assumptions\vspace{2pt}\\
			\hline
			length at the water line  &$L_S=\lambda\, L_M $& geometric similarity\vspace{4pt}\\
			
			wetted surface area of hull&$S_S=\lambda^2\, S_M $&geometric similarity\vspace{4pt}\\
			
			immersed volume&$V_S=\lambda^3\, V_M $&geometric similarity\vspace{4pt}\\
			
			mass of the vessel&$m_{vS}=\lambda^3\, m_{vM}$& the load of the model is prepared in such a way\\&& that the wetted volumes correspond to the \\&&  geometric scaling.\vspace{4pt}\\
			
			hydrodynamic added mass&$m_{addS}=\displaystyle{\lambda^3\,\frac{\rho_S}{\rho_M} m_{addM}}$& 
			the accelerating vessel moves a specific volume \\&&
			of the surrounding water and this volume \\&&scales with respect to geometric similarity.\vspace{4pt}\\
			
			Froude number&${\mathcal Fr}_{S}= {\mathcal Fr}_{M} $&the ratio of the inertia forces to the gravitational\\&& forces related to the mass of water displaced  \\&&by a floating vessel is the same for the model and \\&&full-scale ship.\vspace{4pt}\\
			
			Reinolds number&${\mathcal Re}_{S}=\displaystyle{\lambda^{1.5}\, \frac{\mu_M}{\mu_S}\frac{\rho_S}{\rho_M}{\mathcal Re}_{M}} $& same Froude number and geometric similarity \vspace{4pt}\\
			
			speed&$\textrm{v}_S=\displaystyle{\lambda^{0.5} \,\textrm{v}_M} $&same Froude number and geometric similarity\vspace{4pt}\\
			
			acceleration&$a_S=\lambda^{0.5}\, a_M $&same Froude number and geometric similarity\vspace{4pt}\\
			\hline
			
		\end{tabular}\label{scaling}
	\end{table}
	
	The geometric similarity and the same Froude number for the full-size and model vessels remain the same for the acceleration stage scaling proposition. The difference is that when accelerated, the mass of the vessel and the added mass of water have to be taken into account and scaled. Moreover, since the acceleration is the derivative of speed, for geometric scale $\lambda$, the acceleration scales as
	\begin{equation}
		\frac{\textrm{d} \textrm{v}_S}{\textrm{d} t}=\frac{ \textrm{d} (\lambda^{0.5}\, \textrm{v}_M) }{\textrm{d} t}=\lambda^{0.5}\frac{ \textrm{d} \textrm{v}_M }{\textrm{d} t}.
	\end{equation}

	To derive the scaling formula for the acceleration stage, let us start from Newton's second law of dynamics in the following form
	\begin{eqnarray}	
		m  \,\textrm{v}'
		=
		F_P(\textrm{v},\textrm{v}')-\frac{1}{2}\rho S C_T({\mathcal Fr},{\mathcal Re}) \textrm{v}^2.
	\end{eqnarray}
	Here $\textrm{v} '$ is the speed derivative over time. Further, the breakdown of resistance coefficients (\ref{breakdown}) is assumed, i.e.,
	\begin{eqnarray}
		m \,\textrm{v}'
		=
		F_P(\textrm{v},\textrm{v}')
		-\frac{1}{2}\rho S \left(C_W({\mathcal Fr})+C_V({\mathcal Re})\right) \textrm{v}^2.\label{nl}
	\end{eqnarray}
	The wave resistance coefficient $C_W({\mathcal Fr})$ is assumed to depend only on the Froude number and may be derived from equation (\ref{nl}), that is
	\begin{eqnarray}
		C_W({\mathcal Fr})
		&=&
		\frac{2}{\rho S \textrm{v}^2}F_P(\textrm{v},\textrm{v}') 
		-\frac{2 m}{\rho S \textrm{v}^2 } \,\textrm{v}'
		-C_V({\mathcal Re}).
	\end{eqnarray}
	The Froude number is the same for both: the full-size and the model vessel; therefore, for the scaling procedure, the partial dynamic similarity of the wave resistance coefficient $C_W({\mathcal Fr})$ is used:
	\begin{eqnarray}
		C_W({\mathcal Fr})&=&
		\frac{2}{\rho_M S_M \textrm{v}_M^2}\left(F_{PM}(\textrm{v}_M,\textrm{v}'_M)-m_M\, \textrm{v}'_M\right)-C_{V}({\mathcal Re}_{M}),\label{CwM}\\
		C_W({\mathcal Fr})&=&
		\frac{2}{\rho_S S_S \textrm{v}_S^2}\left(F_{PS}(\textrm{v}_S,\textrm{v}'_S)-m_S\, \textrm{v}'_S\right)-C_{V}({\mathcal Re}_{S}).\label{CWShip}
	\end{eqnarray}
	In (\ref{CwM}) $F_{PM}(\textrm{v}_M,\textrm{v}'_M)$ is the towing force from the acceleration stage towing test on the model vessel. Just one towing test up to the maximal speed is needed to access such information. $F_{PS}(\textrm{v}_S,\textrm{v}'_S)$ in (\ref{CWShip}) is the propulsion force needed to predict ships energy requirement for accelerated motion. Further, it is assumed that the gravitational field $g$ is the same for both the model and the full-scale vessels. Then, one can write the wave-making coefficient $C_W({\mathcal Fr})$ for the full-scale vessel (\ref{CWShip}) using the scaling rules in Table \ref{scaling}:
	\begin{eqnarray}
		C_W({\mathcal Fr})&=&
		\frac{2}{\rho S_M  \textrm{v}_M^2\lambda^3}\left(F_{PS}(\textrm{v}_S,\textrm{v}'_S)-\lambda^{3.5}\left( m_{vM}+\displaystyle{\frac{\rho_S}{\rho_M} m_{addM}}\right)  \textrm{v}'_M\right)-C_{V}\left(\lambda^{1.5}\frac{\mu_M}{\mu_S}\frac{\rho_S}{\rho_M}
		{\mathcal Re}_{M}\right).\label{CwS}
	\end{eqnarray}
	Below let us write an equation where the upper part is the wave-making coefficient $C_W({\mathcal Fr})$ for the full-size vessel with the scaling rules applied (\ref{CwS}), and the bottom part is the wave-making coefficient for the model vessel (\ref{CwM}):
	\begin{eqnarray}
		&\displaystyle{\frac{2}{\rho S_M  \textrm{v}_M^2 \lambda^3}}\left(F_{PS}(\textrm{v}_S,\textrm{v}'_S)- \lambda^{3.5}\left( m_{vM}+\displaystyle{\frac{\rho_S}{\rho_M} m_{addM}}\right) \textrm{v}'_M\right)-C_{V}\left(\lambda^{1.5}\frac{\mu_M}{\mu_S}\frac{\rho_S}{\rho_M}{\mathcal Re}_{M}\right)&\\
		&=&\nonumber\\
		&\displaystyle{\frac{2}{\rho S_M \textrm{v}_M^2}}\left(F_{PM}(\textrm{v}_M,\textrm{v}'_M)-(m_{vM}+m_{addM}) \textrm{v}'_M\right)-C_{V}({\mathcal Re}_{M}).&
		\label{propulsionForceScaling1}
	\end{eqnarray}
	Then, after basics calculations, the following scaling rules for obtaining the propulsion force for the full-scale vessel from the towing experiments on accelerated model vessel are derived:
	\begin{eqnarray}
		\textrm{v}_S &=& \lambda^{0.5}\, \textrm{v}_M,\nonumber\\
		\textrm{v}'_S &=& \lambda^{0.5}\, \textrm{v}'_M,\nonumber\\
		m_S&=&\lambda^3\left( m_{vM}+\displaystyle{\frac{\rho_S}{\rho_M} m_{addM}}\right),\nonumber\\
		F_{PS}(\textrm{v}_S,\textrm{v}'_S)
		&=&
		\lambda^3
		F_{PM}(\textrm{v}_M,\textrm{v}'_M)\label{Fterm}\\
		&+&
		\lambda^3(\lambda^{0.5}-1)m_{vM} \textrm{v}'_M\label{massterm}\\
		&+&
		\lambda^3(\lambda^{0.5}\frac{\rho_S}{\rho_M}-1)m_{addM} \textrm{v}'_M\label{addmassterm}\\
		&+&
		\lambda^3\frac{\rho S_M \textrm{v}_M^2}{2}
		\left(
		C_{V}\left(\lambda^{1.5}\frac{\mu_M}{\mu_S}\frac{\rho_S}{\rho_M}
		{\mathcal Re}_{M}\right)-C_{V}({\mathcal Re}_{M})
		\right).\label{Cvterm}	\label{propulsionForceScaling}
	\end{eqnarray}
	Here, terms (\ref{Fterm}) and (\ref{propulsionForceScaling}) are equivalent to the ITTC procedures for constant speed, when $\textrm{v}'=0$, that is
	\begin{eqnarray}
		&&
		F_{PS}(\textrm{v}_S,0)
		=
		\lambda^3
		F_{PM}(\textrm{v}_M,0)
		+
		\lambda^3\frac{\rho S_M \textrm{v}_M^2}{2}
		\left(
		C_{V}\left(\lambda^{1.5}\frac{\mu_M}{\mu_S}\frac{\rho_S}{\rho_M}\mathcal{R}_{eM}\right)-C_{V}(\mathcal{R}_{eM})
		\right).
		\label{propulsionForceScaling0}
	\end{eqnarray}
	
	Term (\ref{massterm}) is part of the propulsion force needed to accelerate a full-scale vessel. This part is equal to the part of the towing force needed to accelerate the model vessel with corresponding acceleration $\textrm{v}'_M$ times the scaling factor $\lambda^3(\lambda^{0.5}-1)$. Finally, the term (\ref{addmassterm}) is part of the propulsion force that is needed to accelerate the added water mass of the full-scale vessel, and this is equal to the part of the towing force needed to accelerate the added water mass of the model vessel with corresponding acceleration $\textrm{v}'_M$ times the scaling factor $\lambda^3(\lambda^{0.5} \rho_S/\rho_M-1)$. Different water densities for the full-scale and model vessels were considered for the scaling factor in (\ref{addmassterm}). 
	
	Therefore the scaling approach for calculating the propulsion force of the full-scale vessel from towing experiment on the model vessel with accelerated motion is proposed below: 
	\begin{enumerate}[label=Step \arabic*:, leftmargin=*,itemsep=0mm]
		\item Setting the full-scale ship speed $\textrm{v}_S$ range, from the minimum to the desired maximum speed.
		\item Calculating the towing speeds for the model $\textrm{v}_M$ using the Froude's law of similarity (\ref{similarity}).
		\item Recording the towing force $F_{PM}(\textrm{v}_M,\textrm{v}'_M)$ of the model vessel from the acceleration stage towed up to the maximal speed.
		\item Calculating the added mass by extrapolating the towing force to zero speed and using formula (\ref{addM}).
		\item Determining the propulsion force $F_{PS}(\textrm{v}_S,\textrm{v}'_S)$ using formula terms (\ref{Fterm}) - (\ref{propulsionForceScaling}).
	\end{enumerate}

	One remark is that no time scale has been used in the proposed scaling procedure. The equations of motion for any profile can be derived from the second law of dynamics after determining the propulsion force of a full-size vessel $F_{PS}(\textrm{v}_S,\textrm{v}'_S)$. 

	\section*{\centering CONCLUSIONS}\label{sec:}
	\qquad
	In this work, a dynamical scaling proposition for the propulsion force needed to estimate the full-scale vessel energy requirement is derived. The towing force can be experimentally measured from the acceleration stage tests on a model vessel. This theoretical analysis shows that such an approach may have advantages compared with constant speed towing tests. From the acceleration stage, it is possible to obtain information about the hydrodynamic added mass, which should also be taken into account when predicting the ship's energy consumption. Furthermore, in the acceleration stage, all the information about the constant speed stage is accessible. Finally, such towing tests can predict dynamical models that can simulate different motion profiles: constant speed, accelerating, decelerating, and gliding. 
	\bibliographystyle{agsm} 

\begin{thebibliography}{1}
		\bibitem[Birk(2019)]{Birk2019}
		Birk, L. (2019). Fundamentals of ship hydrodynamics: Fluid mechanics, ship resistance and propulsion. John Wiley \& Sons.
		%
		%
		\bibitem[Froude(1872)]{Froude1872}
		Froude, W. (1872). Experiments on the surface-friction experienced by a plane moving through water. British Association for the Advancement of Science, 42, 118-124.
		%
		\bibitem[Froude(1874)]{Froude1874}
		Froude, W. (1874). On experiments with HMS Greyhound. Trans, INA, 15.
		%
		\bibitem[Froude(1877)]{Froude1877}
		Froude, W. (1877). Experiments upon the effect produced on the wave-making resistance of ships by length of parallel middle body. Institution of Naval Architects.
		%
		\bibitem[Ghassemi \& Yari(2011)]{Ghassemi2011}
		Ghassemi, H., \& Yari, E. (2011). The Added Mass Coefficient computation of sphere, ellipsoid and marine propellers using Boundary Element Method. Polish Maritime Research, 18(1), 17-26.
		%
		\bibitem[Havelock(1909)]{Havelock1909}
		Havelock, T. H. (1909). The wave-making resistance of ships: a theoretical and practical analysis. Proceedings of the Royal Society of London. Series A, Containing Papers of a Mathematical and Physical Character, 82(554), 276-300.
		%
		\bibitem[(1932)]{Havelock1932}
		Havelock, T. H. (1932). The theory of wave resistance. Proceedings of the Royal Society of London. Series A, Containing Papers of a Mathematical and Physical Character, 138(835), 339-348.
		%
		\bibitem[Hughes(1954)]{Hughes1954}
		Hughes, G. (1954). Friction and form resistance in turbulent flow and a proposed formulation for use in model and ship correlation. Transactions of the Royal Institution of Naval Architects, 96, 314–376.
		%
		\bibitem[Imlay(1961)]{Imlay1961}
		Imlay, F. H. (1961). The complete expressions for added mass of a rigid body moving in an ideal fluid. David Taylor Model Basin Washington DC.
		%
		\bibitem[Iversen \& Balent(1951)]{IversenBalent1951}
		Iversen, H. W., \& Balent, R. (1951). A correlating modulus for fluid resistance in accelerated motion. Journal of Applied Physics, 22(3), 324-328.
		%
		\bibitem[Kunicka \& Litwin(2019)]{Kunicka2019}
		Kunicka, M., \& Litwin, W. (2019). Energy demand of short-range inland ferry with series hybrid propulsion depending on the navigation strategy. Energies, 12(18), 3499.
		%
		\bibitem[Michell(1898)]{Michell1898}
		Michell, J. H. (1898). XI. The wave-resistance of a ship. The London, Edinburgh, and Dublin Philosophical Magazine and Journal of Science, 45(272), 106-123.
		%
		\bibitem[Motora(1960)]{Motora1960}
		Motora, S. (1960). On the measurement of added mass and added moment of inertia of ships in steering motion. In Proceedings of the First Symposium on Ship Maneuverability, David Taylor Model Basin Report, 1461, 241-274.
		%
		\bibitem[Molland et al.(2017)]{Molland2017}
		Molland, A. F., Turnock, S. R., \& Hudson, D. A. (2017). Ship resistance and propulsion. Cambridge university press.
		%
		\bibitem[Newman(2018)]{Newman2018}
		Newman, J. N. (2018). Marine hydrodynamics. The MIT press.
		%
		\bibitem[Poisson(1832)]{Poisson1832}
		Poisson, S. D. (1832) On the simultaneous movement of a pendulum and of the surrounding air. Mem. Acad. Sci., Paris 11, 521–582.
		%
		%
		\bibitem[Terziev et al.(2022)]{Terziev2022}
		Terziev, M., Tezdogan, T., \& Incecik, A. (2022). Scale effects and full-scale ship hydrodynamics: A review. Ocean Engineering, 245, 110496.
		%
		\bibitem[Kelvin(1887)]{Thomson1887}
		Thomson, W. (1887). On ship waves. Proceedings of the institution of mechanical engineers, 38(1), 409-434.
		%
		\bibitem[Tuck(1989)]{Tuck1989}
		Tuck, E. O. (1989). The wave resistance formula of JH Michell (1898) and its significance to recent research in ship hydrodynamics. The ANZIAM Journal, 30(4), 365-377.
		%
		%
		\bibitem[Zeraatgar et al.(2020)]{Zeraatgar2020}
		Zeraatgar, H., Moghaddas, A., \& Sadati, K. (2020). Analysis of surge added mass of planing hulls by model experiment. Ships and Offshore Structures, 15(3), 310-317.
		%
	\end{thebibliography}
	
\end{document}